# The Quadrantids and December alpha Draconids 2012-2019 - Multi-year Meteor Videography

Alex Pratt

## Abstract

NEMETODE, a network of low-light video cameras in and around the British Isles, operated in conjunction with the BAA Meteor Section and other groups, monitors the activity of meteors, enabling the precision measurement of radiant positions and the altitudes and geocentric velocities of meteoroids and their solar system orbits. The results from observations of the Quadrantid and December alpha Draconid meteor showers during 2012-2019 are presented and discussed.

## Equipment and methods

The NEMETODE team employed the equipment and methods described in previous papers[1],[2] and on their website.[3] These include Genwac, KPF and Watec video cameras equipped with fixed and variable focal length lenses ranging from 2.6mm semi fish-eye models to 12mm narrow field systems.

## The Quadrantid and December alpha Draconid meteor streams

The Quadrantids, IAU Meteor Data Centre code (010 QUA) are medium-speed meteors with geocentric velocities ($V_g$) of 41 km/s, active from late December to mid-January, appearing from a radiant in northern Boötes (the now defunct constellation Quadrans Muralis: the Mural Quadrant). Their radiant is circumpolar from the British Isles, although it is low in the northern skies until after local midnight, climbing to an altitude of about 75° at the onset of morning twilight.

The QUAs produce low rates except for a ZHR (Zenithal Hourly Rate) of ~120 at maximum at solar longitude ($\lambda_\odot$) 283°.16 (2019 January 4th 02h 20m UT) according to the IMO [4], during a brief FWHM (full width half maximum) period of about 14 hours.[5] Their parent body has been identified as the Amor NEO (196256) 2003 $EH_1$, although comet 96P/Machholz 1 has also been proposed as their progenitor. [6,7]

The December alpha Draconids, (334 DAD) were discovered 10 years ago by the Japanese video meteor network, SonotaCo. They described this meteor shower as being active from late November to the end of December with a maximum at $\lambda_\odot$ 256°.5 (2018 December 8) and having a $V_g$ very similar to that of the QUAs.[8] Professional radar studies have shown that both streams are members of a northern toroidal group, a source of meteoroids that is highly inclined to the ecliptic.[9]

In late December the daily motion of the QUA and DAD shower radiants brings them into close proximity with each other and having similar $V_g$ increases the likelihood of single-station meteors being assigned to the wrong stream. Triangulation from 2 or more stations was used in this paper to identify candidates from the QUA and DAD radiants.

## First results

The first probable single-station DAD and QUA meteors in the NEMETODE dataset [10] were recorded on 2010 Nov 16 and 2010 Dec 19, respectively, but it was 2012 Dec before the team obtained multi-station DADs and 2014 Jan for the QUAs. The latter were reported in the 2015 June Journal.[2]  At the 2017 Jan 21 Ordinary Meeting of the Association the author presented an update on multi-year Quadrantid results, including a discussion of their close similarity with the December alpha Draconids.[11] This paper summarises the results from QUA meteors recorded in 2014-2019 Jan and DAD in 2012-2018 Dec.

## Meteor stream catalogues

*UFO Analyser* and *UFO Orbit* utilise a meteor stream catalogue of each shower's start, maximum and end dates, radiant drift and geocentric velocity. *UFO Analyser* assigns a stream category to each single-station meteor and by default it extends the search window by 10 days before and after the normal shower date limits. To reduce the chance of a meteor being incorrectly assigned to the wrong stream, such as in this case with the overlapping activity timelines of the DADs and QUAs, the extension period was set to 0 days and this parameter setting was also adopted in *UFO Orbit* when identifying multi-station QUA and DAD meteors.

The most comprehensive catalogue distributed with SonotaCo's *UFO* software suite is ULE_J6 (known as J6). EDMONd, the European viDeo MeteOr Network [12] has produced a J8 catalogue which includes many of the established minor streams missing from J6. The J8 catalogue is undergoing more frequent revision and version J8_02 (August 2019) was considered when analysing DAD and QUA meteors, with the following caveats.

It extends the QUA start and end limits by 8 days more than J6, further increasing the overlap with the DAD shower in December. If the software also added 10 days to these limits this would give an exceedingly lengthy window of Dec 9 to Jan 29 for a shower with such a narrow FWHM around Jan 3-4. The QUAs would be deeply entangled with the DADs and some meteors would be incorrectly classified.

J6 and J8_02 use the same parameters for the DADs, except that J8_02 adopts a standard radiant circle of $5°$ radius, instead of the large radius of $9°$ used by SonotaCo in J6. Hence, using J8_02 rejects some DADs that are found by J6, so it cannot provide a good comparison with SonotaCo's results; the Japanese group had to cast a wide net to discover and monitor this shower.

In the stream catalogues the values of radiant drift in RA and Dec ($dRA°$ and $dDec°$, in degrees per degree of solar longitude), play an important role in assigning a meteor to a stream category. The software uses the drift rates to compute the positions of the radiants that are active at the time of a meteor and checks if its trail passes within (or close to) the radiant circles. Any significant error in $dRA°$ and/or $dDec°$ could cause a meteor to be incorrectly classified. In J6, SonotaCo gives a $dDec°$ value of +0.17 for the QUAs, but the IAU MDC lists it as -0.25 and in J8_02 it is -0.2.

For these reasons, this paper used the J6 catalogue but with QUA values of +0.5 and -0.25 for $dRA°$ and $dDec°$ respectively.

**Multi-station meteors**

*UFO Analyser* creates an output file of each observer's single-station meteors. These files are collated into annual datasets and processed by *UFO Orbit* to identify and tabulate all simultaneous meteor events (within 3s timing tolerance) recorded by 2 or more video stations. It is also used to create radiant maps, ground tracks and solar system orbits.

*UFO Orbit* supports three built-in Quality Assurance criteria:-
    Q1 – minimum criteria for radiant computation
    Q2 – standard criteria for radiant and velocity computation
    Q3 – criteria for high precision computation

A search of the NEMETODE dataset for Q1-quality multi-station events found 486 QUAs and 243 DADs from the alignments of 1086 and 533 single-station meteors, respectively. Only meteors that were components of a multi-station QUA or DAD were used in the analyses in this paper. Table 1 lists the observers who contributed video meteor data.

*UFO Analyser* estimates the apparent magnitude (m) of each meteor. The magnitude distributions of the QUA and DAD meteors are given in Figure 1.

The mean apparent magnitudes of the QUA and DAD meteors are 0.2 and 0.4 respectively, suggesting that the DADs could be slightly richer in faint meteors. The mean QUA magnitude is half a magnitude brighter than that given in reference 2. This is because multi-station Q-criteria can eliminate some fainter meteors that are otherwise included in single-station analyses.

**Quadrantid maximum**

The short duration Quadrantid maximum occurs in early January with the advantage of long dark nights, but the disadvantage that inclement weather can thwart attempts at recording their peak activity. Also, if maximum occurs when the radiant is at low altitude or during daytime the observer will experience lower rates. Depending on the clarity of the sky, video cameras can record meteors in bright Moonlight, even at Full Moon. The IAU MDC gives a solar longitude of 283°.0 for Quadrantid maximum.[13] In recent years this equates to the following dates and times (UT) and age of the Moon (Table 2).[14]

So for video observers in the British Isles, these years offered different views of the peak activity of the Quadrantids during various lunar phases.

To build a multi-year profile, the 1086 QUAs were corrected for radiant altitude and the average meteor counts per camera were grouped into bands of 1° solar longitude (~24 hours) (Figure 2).

A sharp and distinctive peak at $\lambda_\odot$ 283° emerged from the inherent variability of the meteor shower and the effects of variable sky conditions on our data. To display the peak in more detail, the QUAs were grouped into bands of 0°.1 solar longitude (~2.4 hours) (Figure 3).

This suggests that Quadrantid maximum occurred at $\lambda_\odot$ 283°.0 with a FWHM duration of about 1 day, from around $\lambda_\odot$ 282°.5 to $\lambda_\odot$ 283°.5. This interval is likely to be broader than that observed in any single year due to this profile being derived from combining several years' data.

**December alpha Draconid activity profile**

The 533 DAD meteors were corrected for radiant altitude and the average meteor counts per camera were grouped into bands of 1° solar longitude (Figure 4).

There is no indication of a significant peak in DAD activity and no evidence in this dataset of a maximum near $\lambda_\odot$ 256° as listed by the IAU MDC.[15] However, there is a small rise at $\lambda_\odot$ 261° and the similar increase at $\lambda_\odot$ 273° is caused by a few meteors at low altitudes with large correction factors, otherwise the DADs appear to produce consistently low rates throughout the month of December.

**Radiant Drift**

Because of our changing viewpoint of a meteor stream from the Earth as we orbit the Sun, a meteor shower's radiant slowly moves against the background of stars. *UFO Orbit* was used to derive the radiant point for each Q1 multi-station DAD and QUA meteor (corrected for zenith attraction) and they are plotted in Figures 5 and 6.

There is some overlap of the QUA and DAD radiants at around $\lambda_\odot$ 277° (2018 Dec 28-29). Figure 7 uses multi-year data to illustrate the close proximity of their radiants. Note the diffuse spread of DADs compared with the more compact QUAs.

Interestingly, the radiant plots display a cluster of QUA activity between $\lambda_\odot$ 286° and $\lambda_\odot$ 288° (Jan 6 – 8) when we would expect rates to be in decline. This can also be seen in Figure 2.

The data for Figures 5 and 6 were used to estimate the daily drift in Right Ascension and Declination (per degree of solar longitude) of the QUA and DAD shower radiants.

The method of least squares gave the following linear fits:-

QUA
    RA = (0.462 * $\lambda_\odot$) + 98.741
    Dec = (-0.213 * $\lambda_\odot$) + 110.15

DAD
    RA = (0.459 * $\lambda_\odot$) + 89.865
    Dec = (-0.225 * $\lambda_\odot$) + 117.51

The radiant drift rates of both showers are almost identical.

If we assume that Quadrantid maximum occurred at λ☉ 283°.0, its radiant is then estimated to be located at RA is 229°.5 (15h 18m) and Declination 49°.9. The estimated drift of the radiant per degree of solar longitude (~ per calendar day) is +0°.46 in RA and -0°.21 in Declination. These are presented in Table 3 in comparison with other sources.

**Detection and extinction altitudes**

*UFO Orbit* computed the start and end altitudes of the Q1-quality multi-station QUA and DAD meteors and estimated their apparent magnitudes (m), from which it derived their absolute magnitudes (M) (Figures 8 and 9). (Absolute magnitude is the brightness the meteor would have if it was observed in the zenith, 100km above the observer).

The method of least squares gives the linear fits:

QUA
  Detection altitude (km) = (0.081 * M) + 97.764
  Extinction altitude (km) = (2.938 * M) + 88.082

DAD
  Detection altitude (km) = (-0.019 * M) + 98.221
  Extinction altitude (km) = (2.521 * M) + 87.73

The altitude profiles of the QUA and DAD meteors are almost identical.

**Geocentric velocities**

*UFO Orbit* computed the geocentric velocities (Vg) of 159 QUA and 79 DAD Q2-quality multi-station meteors, which gave the following:-

QUA   Mean   $40.4 \pm 0.2$ km/s
DAD   Mean   $41.7 \pm 0.5$ km/s

These are compared with other sources in Table 4.

**Orbits**

*UFO Orbit* also computed the orbital elements of 74 QUA and 32 DAD Q3-quality meteors. [3 QUA and 10 DAD meteors with large discrepancies in their semi-major axes were excluded]. A summary of the values for each stream is given in Table 5, compared with other sources. [Tisserand's parameter $T_J$ is computed from a body's orbital elements and in this case it is used to compare its orbit with that of Jupiter. A value of $2 > T_J > 3$ is characteristic of Jupiter-family comets.

The QUA and DAD orbits are very similar, apart from their times of perihelion and their descending nodes when they cross the ecliptic.

Figure 10 is a solar system diagram of the Q3-quality orbits, showing how the Earth encounters first the DAD and then the QUA meteor streams. Both are highly inclined to the ecliptic, so that their meteoroids dive down onto the Earth and the meteors appear from high declination radiants.

**Discussion**

Canadian meteor radar observations classify the QUA stream as a member of a toroidal group which includes the DAD stream and others yet to be confirmed.[9] Koseki (Nippon Meteor Society) discusses the importance of correctly identifying shower members and the problems with extending normal shower limits by several days, with specific reference to the QUA and DAD meteors.[17] Koseki calls for cooperative regulation to be applied to the plethora of 'established' minor showers being added to reference catalogues, some of which are derived from alignments of only a few meteors.

Without careful management of the catalogues there will be multiple active radiants on any given night, some in close proximity, hence when back-tracing an observed meteor trail it is likely to align with more than one radiant, increasing the likelihood of incorrect classification by the analysis software. The shower membership of a single-station meteor is thus probabilistic and provisional.

To mitigate the problem of identification of QUA and DAD meteors, this paper used multi-station triangulation of meteors detected between normal shower limits, and an updated value for QUA radiant drift in Declination. The analyses showed that the DADs produce consistently low rates throughout December, whereas the QUAs present a strong short-lived peak of activity in early January at $\lambda_\odot$ 283.0 (2020 Jan 4 04:45 UT). The DADs are perhaps richer in faint meteors with slightly higher geocentric velocities (Vg), otherwise both streams display the same radiant drift, detection and extinction altitudes, and their orbital characteristics are very similar.

The Quadrantids are the first major shower of the year and it is always a challenge to observe their elusive peak around Jan 3-4. Radio and video studies show they are not an isolated shower; we have a complex story of activity from more than one stream. It will be interesting to see the results from our dataset over the next few years.

### Acknowledgements

The author would like to thank all observers who contributed video meteor data to the NEMETODE dataset, Richard Kacerak for supplying a copy of the J8_02 stream catalogue, Tracie Heywood for bringing Koseki's paper to his attention, and William Stewart for his Tisserand's parameter calculator and for his constructive comments on a first draft of this paper.

**Address**: c/o British Astronomical Association, Burlington House, Piccadilly, London W1J 0DU.

science@nemetode.org

16. http://ssd.jpl.nasa.gov/sbdb.cgi
17. Koseki M, 'Different definitions make a meteor shower distorted. The views from SonotaCo net and CAMS.', *WGN*, 46:4, 119-135 (2018)

**Tables and Figures**

| Observer | Location |
|---|---|
| David Anderson | Low Craighead & Dunure, Scotland |
| Steve Bosley | Clanfield, England |
| Denis Buczynski | Tarbatness, Scotland |
| Peter Carson | Leigh-on-Sea, England |
| Allan Carter | Basingstoke, England |
| Crayford Manor House | Dartford, England |
| David Dunn | Livarot, France |
| Dunsink Observatory | Dunsink, Ireland |
| Mike Foylan | Rathmolyon, Ireland |
| Nick James | Chelmsford, England |
| Frank Johns | Newquay, England |
| Steve Johnston | Warrington, England |
| Jon Jones | Huntington, England |
| Charlie McCormack | Galway, Ireland |
| Andy McCrea | Bangor, Northern Ireland |
| Michael Morris | Worcester, England |
| Michael O'Connell | Monasterevin, Ireland |
| Alex Pratt | Leeds, England |
| Nick Quinn | Steyning, England |
| Gordon Reineke | Newbridge, Ireland |
| Graham Roche | Dublin, Ireland |
| Jim Rowe | East Barnet, England |
| Nick Rowell | Gargunnock, Scotland |
| Jeremy Shears | Bunbury, England |
| Fred Stevenson | Amersham, England |
| Peter Stewart | Derriaghy, Northern Ireland |
| William Stewart | Ravensmoor, England |
| Ray Taylor | Skirlaugh, England |

Table 1 – Observers who contributed video meteor data

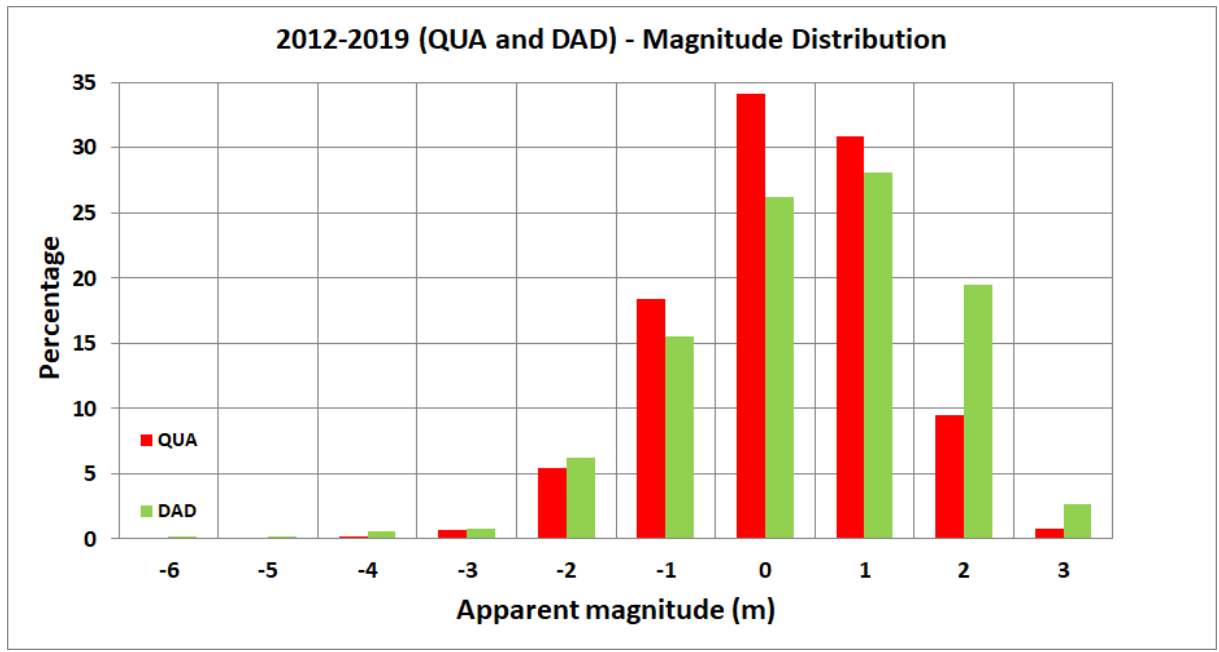

Figure 1 – Magnitude distributions of the QUA and DAD meteors

| QUA maximum | | Moon | |
|---|---|---|---|
| Date and time | Age (days) | Percent illuminated | Radiant distance (degrees) |
| 2014 Jan 3 15:52 | 2 | 7+ | 96 |
| 2015 Jan 3 22:04 | 13 | 98+ | 104 |
| 2016 Jan 4 04:11 | 24 | 32- | 62 |
| 2017 Jan 3 10:15 | 5 | 25+ | 112 |
| 2018 Jan 3 16:25 | 17 | 96- | 84 |
| 2019 Jan 3 22:35 | 28 | 4- | 74 |

Table 2 – Times of Quadrantid maxima and the degree of Moonlight

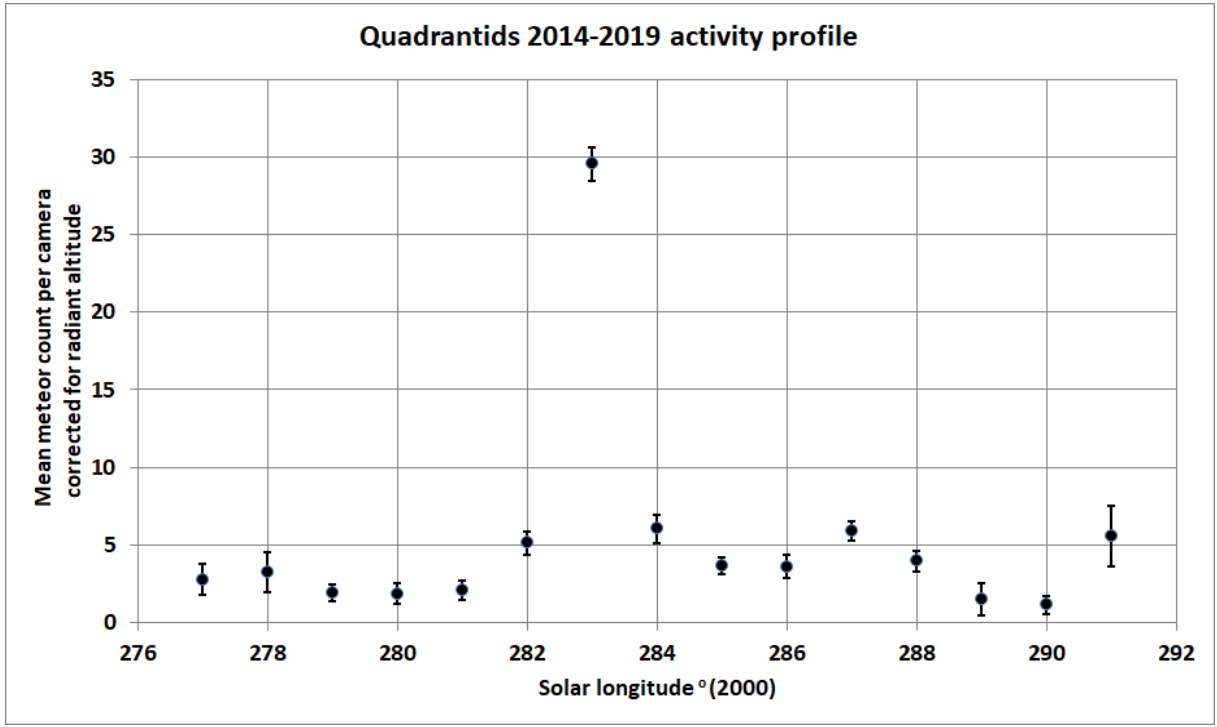

Figure 2 - Quadrantids 2014-2019 activity profile

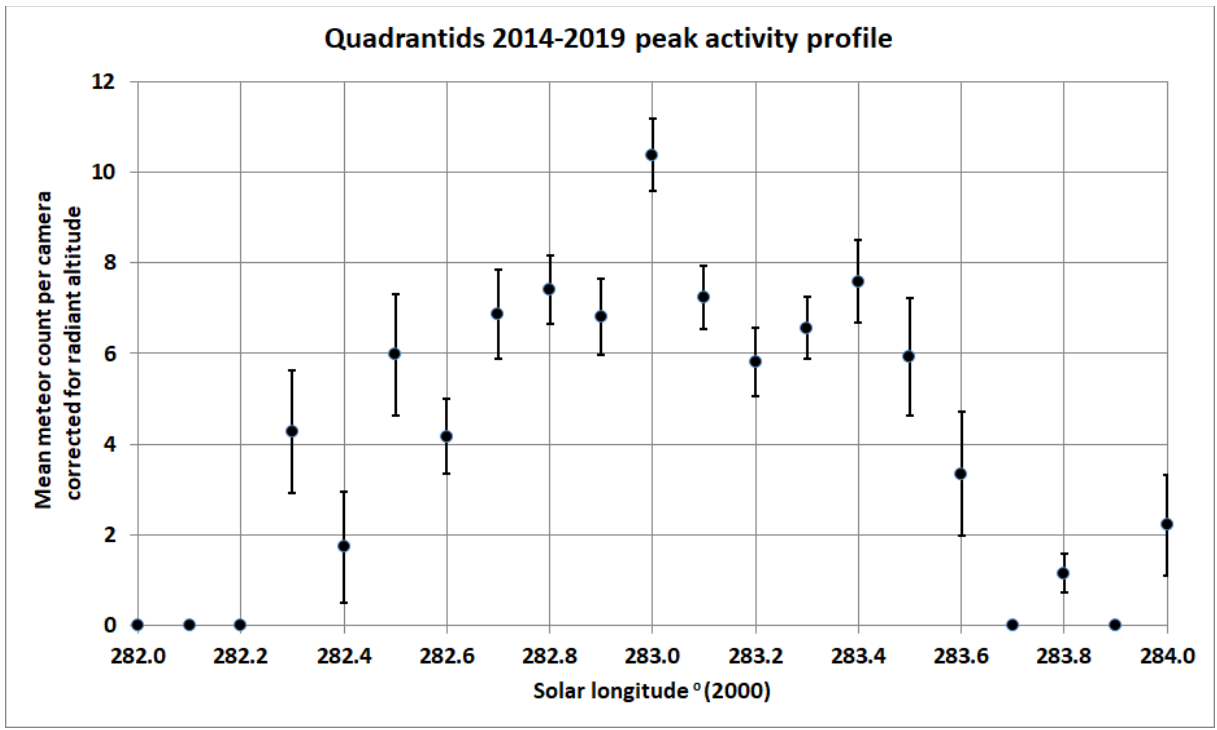

Figure 3 - Quadrantids 2014-2019 peak activity profile

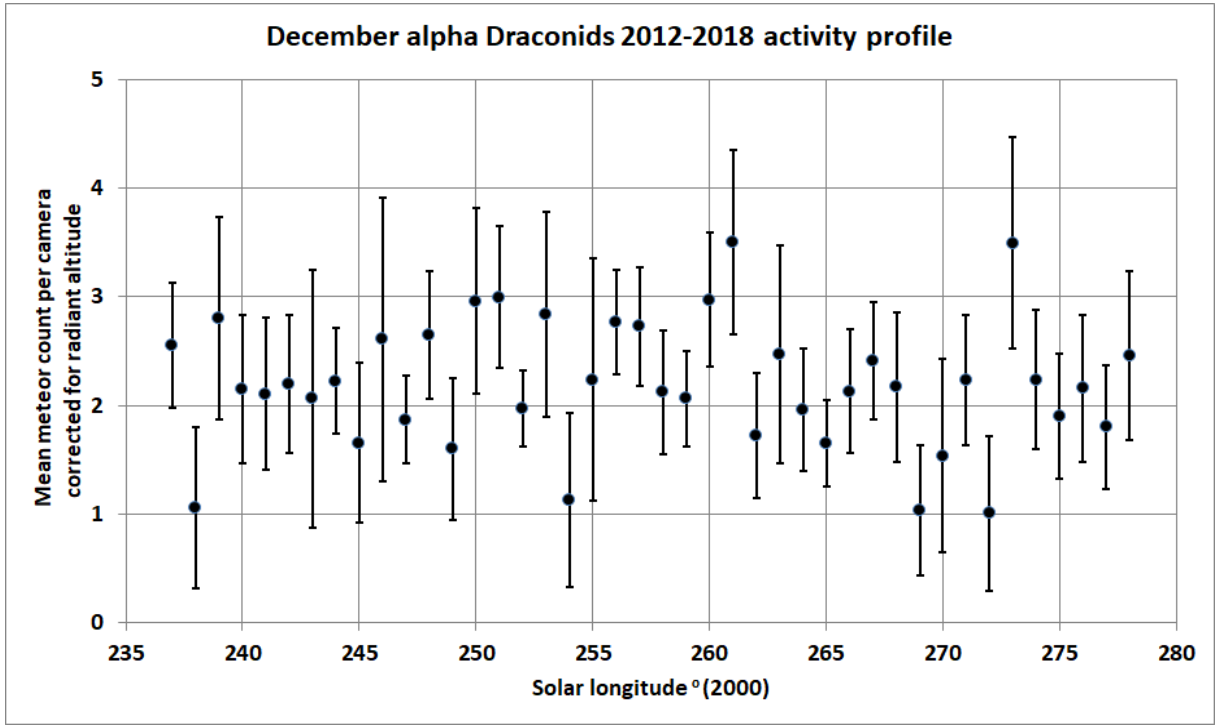

Figure 4 – December alpha Draconids 2012-2018 activity profile

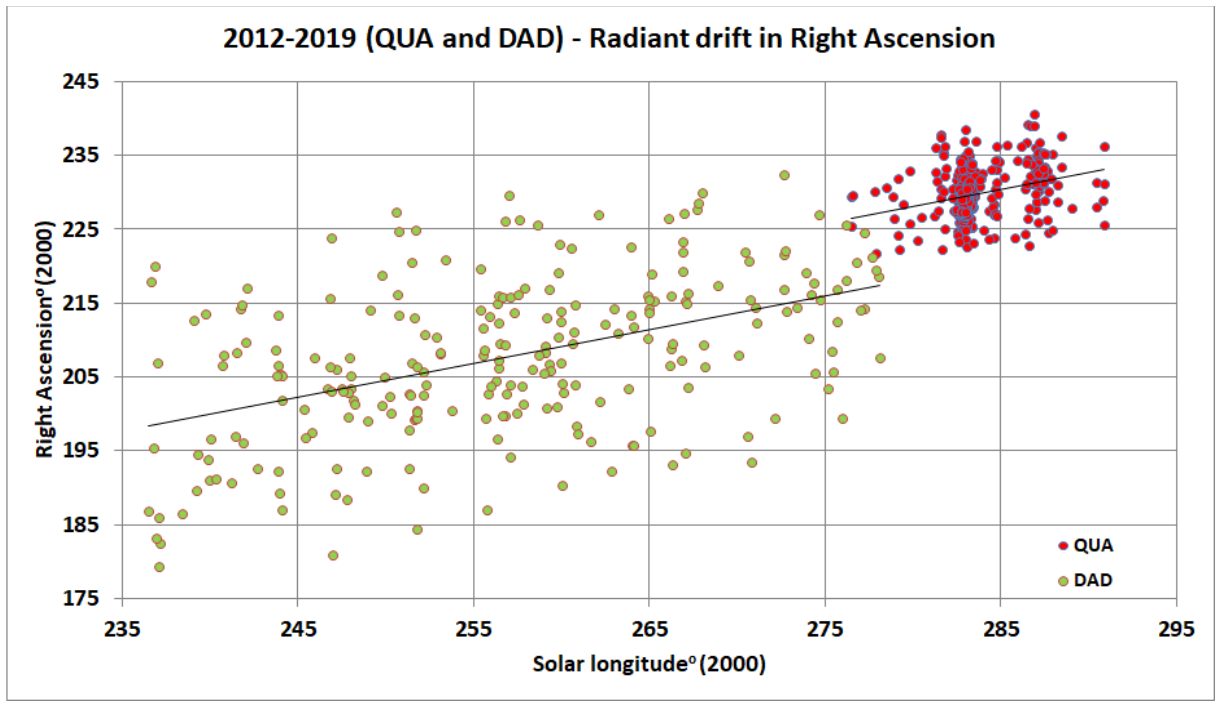

Figure 5 – QUA and DAD radiant drift in Right Ascension

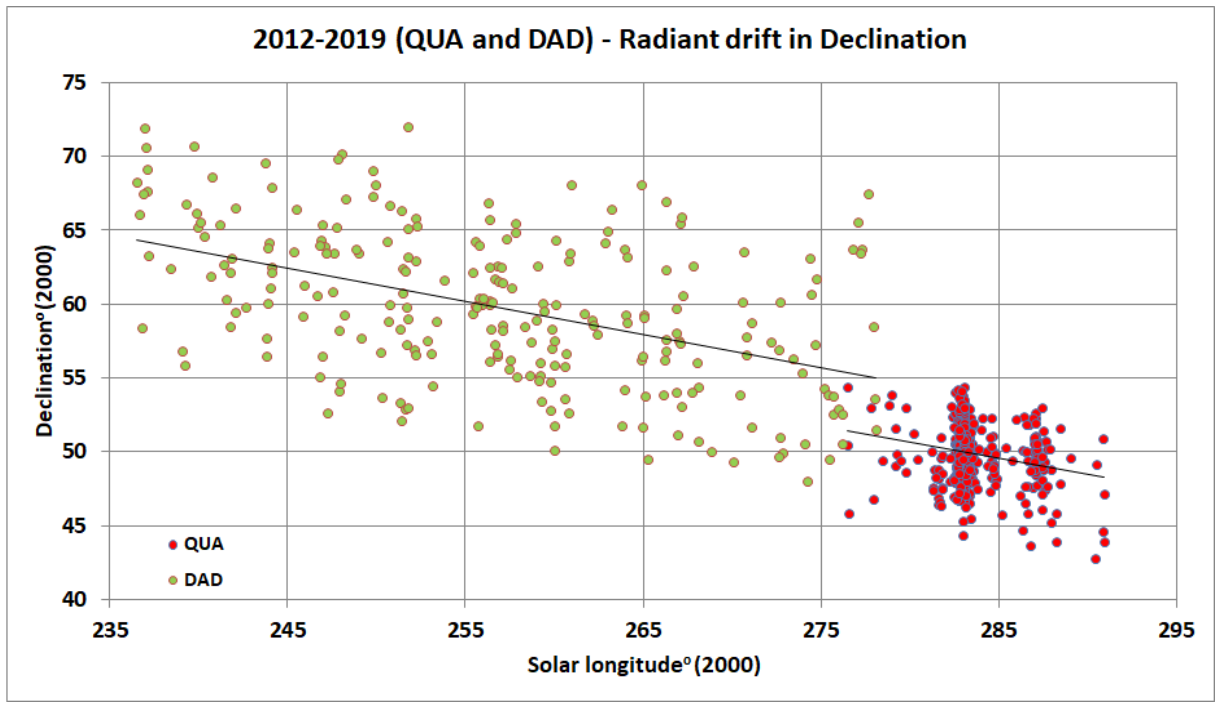

Figure 6 – QUA and DAD radiant drift in Declination

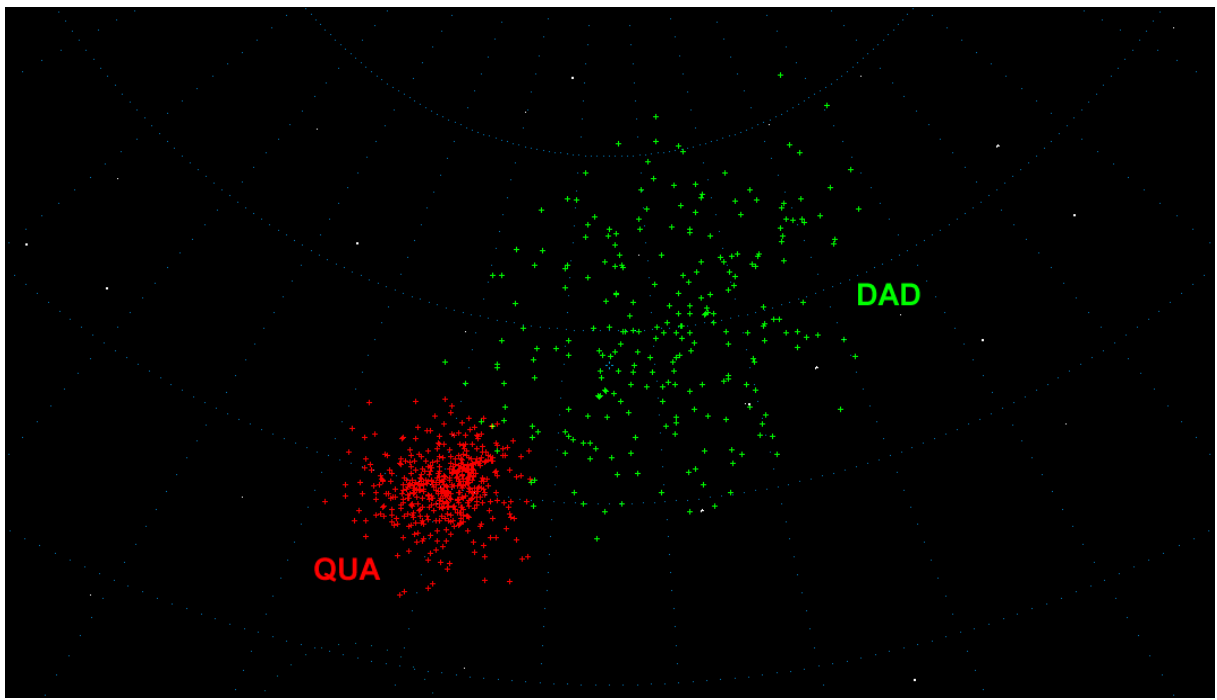

Figure 7 – QUA and DAD radiants

|  | Solar long. ° | RA ° | RA | dRA° | Dec° | dDec° |
|---|---|---|---|---|---|---|
|  | (All positions are for epoch 2000.0) | | | | | |
| **(010 QUA)** | | | | | | |
| IMO [4] | 283.16 | 230 | 15h 20m | 0.6 | 49 | -0.2 |
| Jenniskens *et al* [13] | 283.0 | 230.2 | 15h 21m | 0.56 | 49.5 | -0.25 |
| NEMETODE | 283.0 | 229.5 | 15h 18m | 0.46 | 49.9 | -0.21 |
| SonotaCo [8] | 283.1 | 230.0 | 15h 20m | 0.15 | 49.0 | 0.17 |
|  | | | | | | |
| **(334 DAD)** | | | | | | |
| Jenniskens *et a*l [15] | 256.0 | 210.8 | 14h 03m | 0.58 | 58.6 | -0.34 |
| NEMETODE |  |  |  | 0.46 |  | -0.23 |
| SonotaCo [8] | 256.5 | 207.9 | 13h 52m | 0.40 | 60.6 | -0.14 |

Table 3 – The position of the Quadrantid radiant at maximum, its daily motion and DAD data

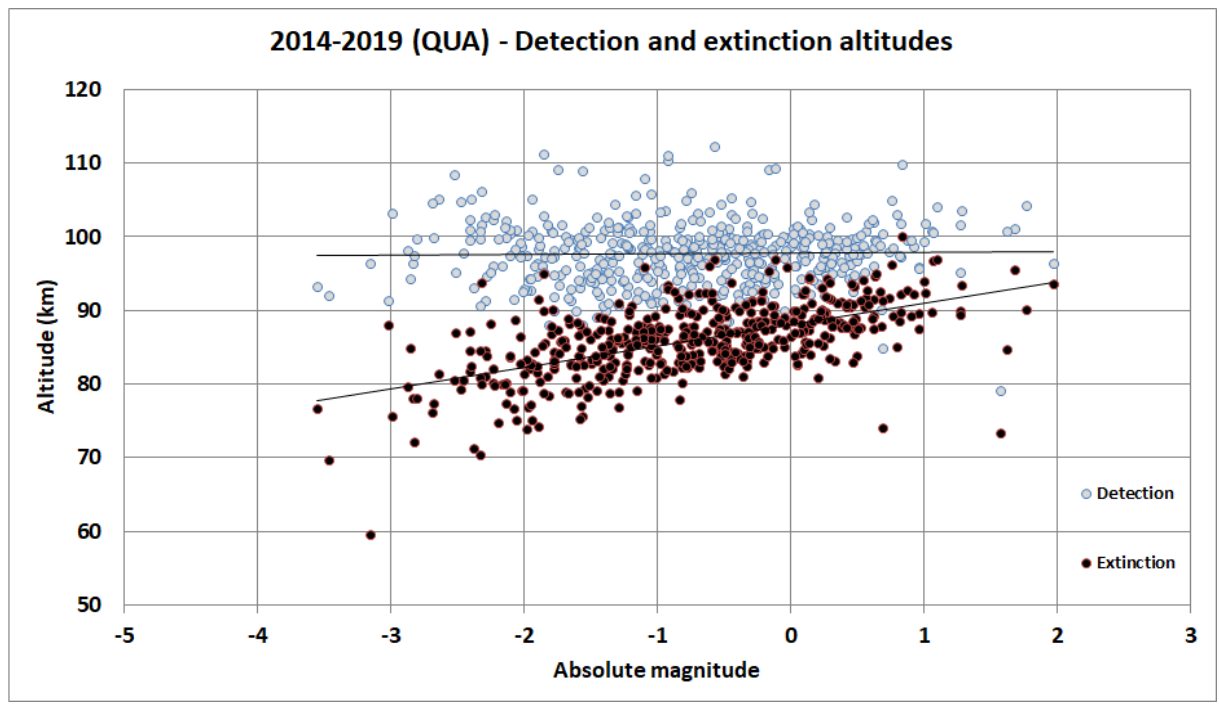

Figure 8 – Detection and extinction altitudes of QUA meteors

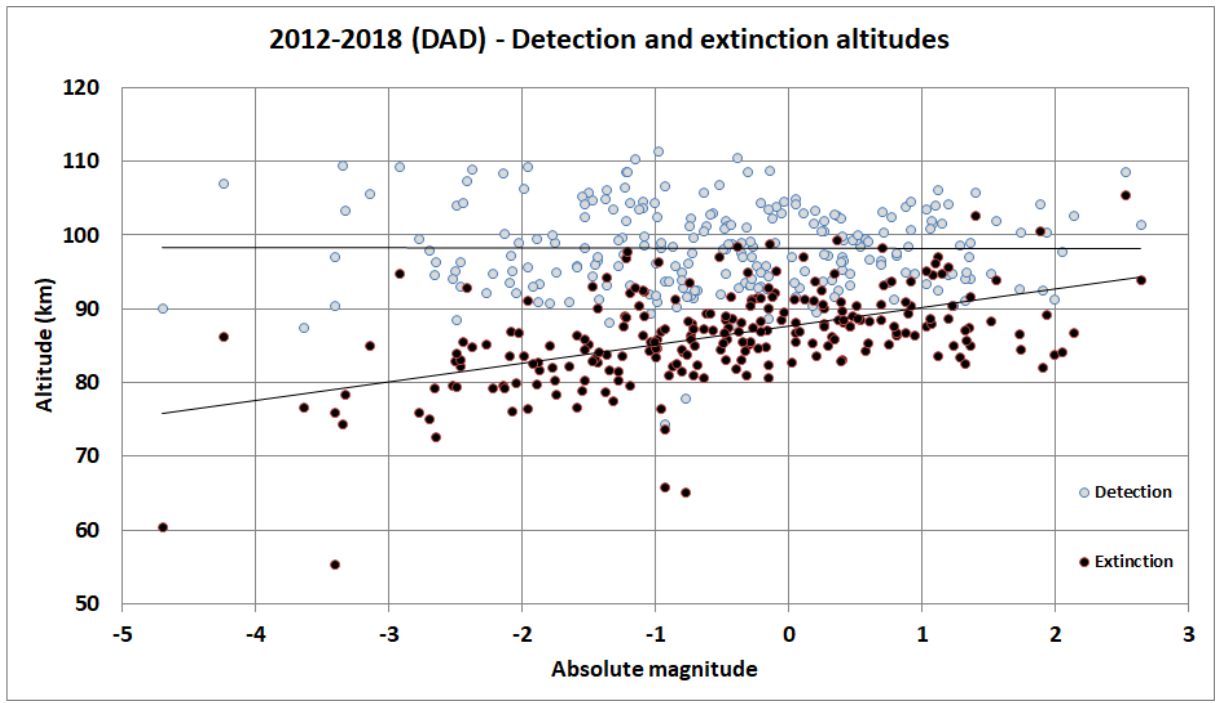

Figure 9 – Detection and extinction altitudes of DAD meteors

|  | Vg (km/s) | ± | n |  |
|---|---|---|---|---|
|  |  |  |  |  |
| **(010 QUA)** |  | ± |  |  |
| Brown *et al* [13] | 41.7 |  | 6614 | radar |
| Jenniskens *et al* [13] | 40.7 |  | 1029 |  |
| NEMETODE | 40.4 | 0.2 | 159 |  |
| SonotaCo [9] | 40.0 |  | 243 |  |
|  |  |  |  |  |
| **(334 DAD)** |  | ± |  |  |
| Jenniskens *et al* [15] | 40.8 |  | 47 |  |
| NEMETODE | 41.7 | 0.5 | 79 |  |
| SonotaCo [9] | 41.6 |  | 145 |  |

Table 4 – Geocentric velocities (Vg) of QUA and DAD meteors

|  | a | | q | | e | | Peri | | Node | | Incl | | n | Tisserand's |
|---|---|---|---|---|---|---|---|---|---|---|---|---|---|---|
|  | (AU) | | (AU) | | | | J2000 | | J2000 | | J2000 | | | parameter |
|  | | | | | | | | | | | | | | $T_J$ |
| **(010 QUA)** | | | | | | | | | | | | | | |
|  | | ± | | ± | | ± | | ± | | ± | | ± | | |
| Brown et al [13] | 3.35 | | 0.9746 | | 0.709 | | 168.14 | | 283.0 | | 72.4 | | 6614 | 2.18 |
| Jenniskens et al [13] | 2.82 | | 0.979 | | 0.657 | | 171.4 | | 283.3 | | 71.2 | | 1029 | 2.48 |
| NEMETODE | 2.740 | 0.420 | 0.980 | 0.001 | 0.634 | 0.007 | 173.266 | 0.520 | 283.882 | 0.229 | 70.711 | 0.272 | 74 | 2.54 |
|  | | | | | | | | | | | | | | |
| (196256) 2003 EH1 [16] | 3.124 | | 1.190 | | 0.619 | | 171.35 | | 282.981 | | 70.840 | | | 2.36 |
| 96P/Machholz 1 [16] | 3.033 | | 0.124 | | 0.959 | | 14.746 | | 94.351 | | 58.539 | | | 2.03 |
|  | | | | | | | | | | | | | | |
| **(334 DAD)** | | | | | | | | | | | | | | |
|  | | ± | | ± | | ± | | ± | | ± | | ± | | |
| Jenniskens et al [15] | 2.48 | | 0.983 | | 0.603 | | 177.4 | | 254.8 | | 71.8 | | 47 | 2.71 |
| NEMETODE | 2.729 | 0.130 | 0.976 | 0.002 | 0.618 | 0.019 | 182.685 | 2.308 | 254.193 | 1.866 | 71.010 | 1.220 | 32 | 2.56 |

Table 5 – Orbital elements of QUA and DAD meteors

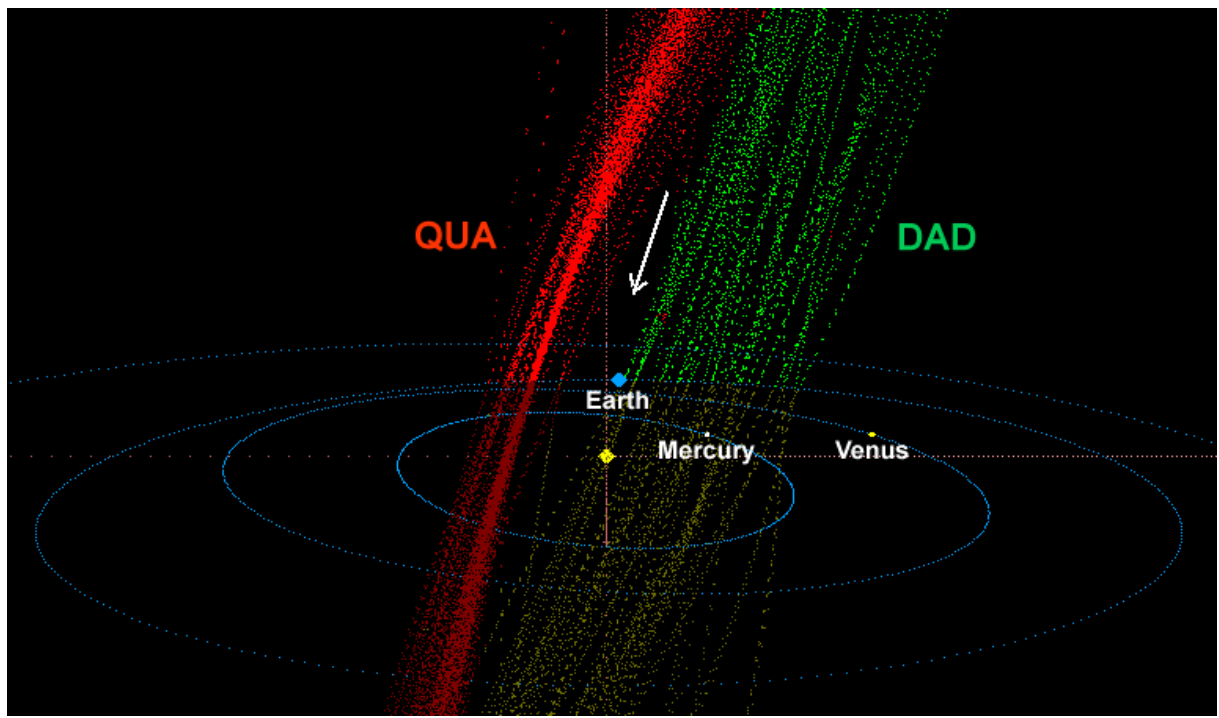

Figure 10 – Solar system orbits of QUA and DAD meteors